\documentclass[titlepage,12pt]{amsart}
 \usepackage{amsaddr}
\usepackage{amsmath, amsthm}              
\usepackage{graphicx}
\usepackage{amssymb}
\usepackage{amsthm, wasysym,color}
\usepackage[all]{xy}
\usepackage{tikz}
\usepackage{url}
\usepackage{colortbl,multirow}
\usepackage{hhline}

\usetikzlibrary{shapes}

\newtheorem{theorem}{Theorem}[section]
\newtheorem{remark}[theorem]{Remark}

\makeatletter
\g@addto@macro{\endabstract}{\@setabstract}
\newcommand{\authorfootnotes}{\renewcommand\thefootnote{\@fnsymbol\c@footnote}}%
\makeatother

\begin{document}

\begin{center}
  \LARGE 
  Counting chemical reaction networks with NAUTY \par \bigskip

  \normalsize
  \authorfootnotes
  Murad Banaji\footnote{Department of Design Engineering and Mathematics, Middlesex University, London. m.banaji@mdx.ac.uk}\par \bigskip


\today
\end{center}

\begin{abstract}
It is useful to have complete lists of nonisomorphic chemical reaction networks (CRNs) of a given size, with or without various restrictions. One may, for example, be interested in exploring how often certain dynamical behaviours occur in small CRNs, or wish to find examples to illustrate some aspect of the theory. In such cases, it is natural to examine one representative from each isomorphism class of CRNs. Inspired by the related project of Deckard {\em et al} \cite{deckard}, this document outlines the methodology involved in listing all CRNs in various classes of interest including, for example, general CRNs, dynamically nontrivial CRNs, weakly reversible CRNs, fully open CRNs, etc. The accompanying data (i.e., lists of nonisomorphic CRNs in the various classes) is at \url{https://reaction-networks.net/networks/}. Note that both document and data are work in progress.
\end{abstract}


\section{Introduction}

When discussing the combinatorial structure of a chemical reaction network (CRN), the basic objects are the chemical {\em species}, the {\em complexes} and the {\em reactions}. Formally, the chemical species of a CRN are an abstract finite set $\mathcal{S}$, the complexes are formal linear combinations of elements of $\mathcal{S}$, namely each complex is a function $F\colon \mathcal{S} \to \mathbb{Z}_{\geq 0}$, and the reactions are ordered pairs of complexes. 

Suppose that the species of a CRN are $\mathrm{X}_1, \ldots, \mathrm{X}_k$. Each complex is of the form $a_1\mathrm{X}_1 + a_2\mathrm{X}_2 + \cdots + a_k\mathrm{X}_k$ where each $a_i$ is a nonnegative integer, the {\em stoichiometry} of $\mathrm{X}_i$ in the complex. A complex $a_1\mathrm{X}_1 + a_2\mathrm{X}_2 + \cdots + a_k\mathrm{X}_k$ satisfying $\sum_ia_i \leq 2$ is termed an {\em at most bimolecular} complex, or a {\bf 2-complex} for short. The {\em zero complex} $0\mathrm{X}_1 + \cdots + 0\mathrm{X}_k$ is denoted $0$ (note that $0$ is a $2$-complex). By basic principles of choice with repetition, there are 
\[
n_C(k):={k+2 \choose 2}
\]
2-complexes on $k$ species, including the zero complex. 

An ireversible reaction is an ordered pair of complexes, the {\em source complex} (or left hand side of the reaction) and the {\em target complex} (or right hand side of the reaction). Thus any chemical reaction on the set of species $\{\mathrm{X}_i\}$ takes the form
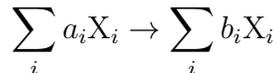
\[
\sum_ia_i\mathrm{X}_i \rightarrow \sum_ib_i\mathrm{X}_i
\]
Such a reaction is at most bimolecular if its source and target complexes are 2-complexes. From here on, at most bimolecular CRNs will be termed {\bf 2-CRNs}, and only 2-CRNs are treated; the modifications required if larger stoichiometries are allowed are fairly straightforward. Two common conventions about CRNs adopted here are:
\begin{enumerate}
\item The source and target complexes of a reaction are distinct;
\item Two distinct reactions cannot have the same source and target complexes.
\end{enumerate}
The first of these conventions is quite natural from the point of view of modelling the dynamics, although there are occasions where we might want to dispense with it for technical reasons. The second is somewhat arbitrary: with certain choices of kinetics, allowing the ``same reaction'' to figure more than once in a  CRN can enlarge the set of allowed models of the CRN, while for others (most importantly, mass action) it cannot. 

\subsection{The complex graph} A CRN is naturally identified with its {\em complex graph}, the digraph whose vertices are the complexes of the network and whose arcs are its (irreversible) reactions. The two conventions mentioned above mean that the complex graph of a CRN must be simple: it has no loops or parallel arcs (antiparallel arcs are of course allowed, corresponding to reversible reactions). Equivalently, its adjacency matrix is a $(0,1)$-matrix with zeros on the diagonal. The complex graph is a core object of classical chemical reaction network theory, and in particular, deficiency theory \cite{feinberg}. 

\subsection{Petri net graphs and isomorphism} For the purposes of discussing isomorphism, the most useful representation of a CRN is not its complex graph, but rather its {\em Petri net (PN) graph} \cite{angelipetrinet}, an edge-weighted bipartite digraph (equivalently, since edge-weights in this case are positive integers, a bipartite multidigraph). The PN graph of a CRN $\mathcal{R}$, denoted $PN(\mathcal{R})$, has two vertex sets $V_S$ (species vertices) and $V_R$ (reaction vertices) identified with the species and reactions of $\mathcal{R}$. Given $\mathrm{X}_i \in V_S$ and $R_j \in V_R$, there exists an arc $\mathrm{X}_iR_j$ (resp., $R_j\mathrm{X}_i$) with weight $w$ if and only if the species $\mathrm{X}_i$ occurs with stoichiometry $w$ in the source (resp., target) complex of the reaction $R_j$. Consider, for example, the CRN $\mathrm{X}+\mathrm{Y} \rightarrow 2\mathrm{Y}$, $\mathrm{Y} \rightarrow \mathrm{X} \rightleftharpoons 0$. Below are three versions of its PN graph:
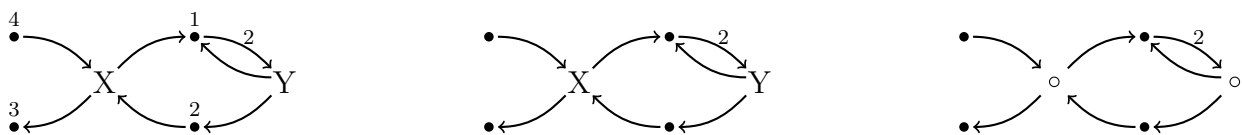
\begin{center}
\begin{tikzpicture}[scale=1.2]
\fill[color=black] (1,0.5) circle (1.5pt);
\fill[color=black] (1,-0.5) circle (1.5pt);
\fill[color=black] (3,0.5) circle (1.5pt);
\fill[color=black] (3,-0.5) circle (1.5pt);

\node at (2,0) {$\mathrm{X}$};
\node at (4,0) {$\mathrm{Y}$};
\node at (1,0.7) {$\scriptstyle{4}$};
\node at (1,-0.3) {$\scriptstyle{3}$};
\node at (3,0.7) {$\scriptstyle{1}$};
\node at (3,-0.3) {$\scriptstyle{2}$};

\draw [<-, thick] (1.85,0.15) .. controls (1.7,0.3) and (1.5,0.5) .. (1.1,0.5);
\draw [->, thick] (1.85,-0.15) .. controls (1.7,-0.3) and (1.5,-0.5) .. (1.1,-0.5);
\draw [->, thick] (2.15,0.15) .. controls (2.3,0.3) and (2.5,0.5) .. (2.9,0.5);
\draw [<-, thick] (3.85,0.15) .. controls (3.7,0.3) and (3.5,0.5) .. (3.1,0.5);
\draw [<-, thick] (2.15,-0.15) .. controls (2.3,-0.3) and (2.5,-0.5) .. (2.9,-0.5);
\draw [->, thick] (3.85,-0.15) .. controls (3.7,-0.3) and (3.5,-0.5) .. (3.1,-0.5);
\draw[->, thick] (3.85,0.05) .. controls (3.5, 0.05) and (3.3, 0.2) .. (3.07,0.43);

\node at (3.6,0.5) {$\scriptstyle{2}$};
\end{tikzpicture}
\hfill
\begin{tikzpicture}[scale=1.2]
\fill[color=black] (1,0.5) circle (1.5pt);
\fill[color=black] (1,-0.5) circle (1.5pt);
\fill[color=black] (3,0.5) circle (1.5pt);
\fill[color=black] (3,-0.5) circle (1.5pt);

\node at (2,0) {$\mathrm{X}$};
\node at (4,0) {$\mathrm{Y}$};

\draw [<-, thick] (1.85,0.15) .. controls (1.7,0.3) and (1.5,0.5) .. (1.1,0.5);
\draw [->, thick] (1.85,-0.15) .. controls (1.7,-0.3) and (1.5,-0.5) .. (1.1,-0.5);
\draw [->, thick] (2.15,0.15) .. controls (2.3,0.3) and (2.5,0.5) .. (2.9,0.5);
\draw [<-, thick] (3.85,0.15) .. controls (3.7,0.3) and (3.5,0.5) .. (3.1,0.5);
\draw [<-, thick] (2.15,-0.15) .. controls (2.3,-0.3) and (2.5,-0.5) .. (2.9,-0.5);
\draw [->, thick] (3.85,-0.15) .. controls (3.7,-0.3) and (3.5,-0.5) .. (3.1,-0.5);
\draw[->, thick] (3.85,0.05) .. controls (3.5, 0.05) and (3.3, 0.2) .. (3.07,0.43);

\node at (3.6,0.5) {$\scriptstyle{2}$};
\end{tikzpicture}
\hfill
\begin{tikzpicture}[scale=1.2]
\fill[color=black] (1,0.5) circle (1.5pt);
\fill[color=black] (1,-0.5) circle (1.5pt);
\fill[color=black] (3,0.5) circle (1.5pt);
\fill[color=black] (3,-0.5) circle (1.5pt);

\draw[color=black] (2,0) circle (1.5pt);
\draw[color=black] (4,0) circle (1.5pt);

\draw [<-, thick] (1.85,0.15) .. controls (1.7,0.3) and (1.5,0.5) .. (1.1,0.5);
\draw [->, thick] (1.85,-0.15) .. controls (1.7,-0.3) and (1.5,-0.5) .. (1.1,-0.5);
\draw [->, thick] (2.15,0.15) .. controls (2.3,0.3) and (2.5,0.5) .. (2.9,0.5);
\draw [<-, thick] (3.85,0.15) .. controls (3.7,0.3) and (3.5,0.5) .. (3.1,0.5);
\draw [<-, thick] (2.15,-0.15) .. controls (2.3,-0.3) and (2.5,-0.5) .. (2.9,-0.5);
\draw [->, thick] (3.85,-0.15) .. controls (3.7,-0.3) and (3.5,-0.5) .. (3.1,-0.5);
\draw[->, thick] (3.85,0.05) .. controls (3.5, 0.05) and (3.3, 0.2) .. (3.07,0.43);

\node at (3.6,0.5) {$\scriptstyle{2}$};
\end{tikzpicture}
\end{center}
In all three, arc-weights of $1$ have been omitted. On the left is the PN graph of a {\em fully labelled} CRN: both species and reaction vertices are labelled; in the middle is the PN graph of a {\em species-labelled} CRN: only species vertices are labelled; and on the right is the PN graph of an {\em unlabelled} CRN (or a {\em motif}) with species-vertices represented as unfilled circles. Given a fully labelled CRN we may relabel species, relabel reactions, or relabel both, but without swapping species and reaction vertices, to get another CRN which is ``fundamentally the same''. According with this intuition, two fully labelled CRNs $\mathcal{R}_1$ and $\mathcal{R}_2$ are {\em isomorphic} if there exists a relabelling of the vertices of $PN(\mathcal{R}_1)$ which preserves the bipartition and gives $PN(\mathcal{R}_2)$. Similarly, two species-labelled CRNs are isomorphic if there is a relabelling of the species which takes one to the other. 

\begin{remark}[Species-labelled CRNs]
A species-labelled CRN can be regarded as an equivalence class of fully labelled CRNs under reaction relabelling. Most authors, when referring to a CRN, implicitly mean a species-labelled CRN, as the order/identity of reactions is not taken to be important: for example (i) $X \rightarrow Y, \,\, 2Y \rightarrow X+Y$ and (ii) $2Y \rightarrow X+Y,\,\,X \rightarrow Y$ are considered the same CRN. From now on, ``chemical reaction network'' without any qualification will refer to a species-labelled CRN. (However, it is worth noting that for some purposes, such as when discussing subnetworks of a CRN, or writing down stoichiometric matrices, it can be important to give the reactions fixed identities.)
\end{remark}

An unlabelled CRN can be regarded as an equivalence class of either fully labelled or species-labelled CRNs under isomorphism. This is made precise in the next section. 

\subsection{Automorphisms of CRNs} 
\label{secautomorph}
Consider the set $\mathcal{R}_{k, l}$ of (species-labelled) 2-CRNs with $k$ species and $l$ reactions on some arbitrary but fixed set of species, and the set $\overline{\mathcal{R}_{k, l}}$ of unlabelled 2-CRNs with $k$ species and $l$ reactions. There is a natural action of the symmetric group $S_k$ on $\mathcal{R}_{k, l}$: elements of $S_k$ permute the species any given CRN in $\mathcal{R}_{k, l}$ to give another CRN in $\mathcal{R}_{k, l}$. The orbit of any CRN under this action has size at most $|S_k| = k!$ and thus each CRN on $k$ species lies in an equivalence class consisting of no more than $k!$ isomorphic CRNs. For example the CRN $\mathcal{R}$: $A+B \rightarrow C,\,\,B \rightarrow 2A$ lies in an equivalence class consisting of $6$ CRNs corresponding to the six permutations of $\{A, B, C\}$, namely:
\[
\begin{array}{l}
\mbox{(i) }A+B \rightarrow C,\,B \rightarrow 2A \quad \mbox{(ii) }A+B \rightarrow C,\,A \rightarrow 2B \quad \mbox{(iii) }B+C \rightarrow A,\,B \rightarrow 2C\\\mbox{(iv) }A+C \rightarrow B,\,C \rightarrow 2A \quad \mbox{(v) }B+C \rightarrow A,\,C \rightarrow 2B \quad \mbox{(vi) }A+C \rightarrow B,\,A \rightarrow 2C\,.
\end{array}
\]
This equivalence class is precisely an element of $\overline{\mathcal{R}_{k, l}}$, namely it is the unlabelled CRN corresponding to $\mathcal{R}$. Define $N_{k,l} = |\mathcal{R}_{k, l}|$ to be the total number of $k$-species, $l$-reactions (species-labelled) 2-CRNs on some fixed set of species (this is evaluated explicitly in Section~\ref{seccountbasic} below). As the orbits of $S_k$ partition $\mathcal{R}_{k,l}$, we get the following lower bound on $\overline{N_{k,l}}$, the number of $k$-species, $l$-reaction unlabelled CRNs, namely
\[
\overline{N_{k,l}} \geq \lceil N_{k,l}/k!\rceil
\]
(here $\lceil \cdot \rceil$ is the ceiling function). The actual size of the isomorphism class of a CRN is less than $k!$ if and only if it has nontrivial symmetries (i.e., nontrivial automorphism group). For example the CRN $A \rightarrow B \rightarrow C \rightarrow A$ has automorphism group isomorphic to $\mathbb{Z}_3$ and consequently there are precisely $6/3=2$ distinct CRNs in its isomorphism class, namely 
\[
\mbox{(i) } A \rightarrow B \rightarrow C \rightarrow A \quad \mbox{and} \quad \mbox{(ii) } A \rightarrow C \rightarrow B \rightarrow A\,.
\] 
Thus, roughly, the difference between $\overline{N_{k,l}}$ and $\lceil N_{k,l}/k!\rceil$ reflects how commonly CRNs with symmetries occur in $\mathcal{R}_{k,l}$ (see data in Table~\ref{tabdat} later).

\subsection{ODE models}
\label{secode}
In order to describe some interesting classes of CRNs we need brief discussion of differential equation models of CRNs. 

A real vector $x = (x_1, \ldots, x_k)^t$ is nonnegative (resp., positive) if $x_i\geq 0$ (resp., $x_i > 0$) for each $i$. The set of nonnegative (resp., positive) vectors in $\mathbb{R}^k$ is denoted $\mathbb{R}^k_{\geq 0}$ (resp., $\mathbb{R}^k_{\gg 0}$). Subsets of $\mathbb{R}^k_{\gg 0}$ are referred to as positive. $x \gg 0$ means $x_i > 0$ for each $i$; $x \geq 0$ means $x_i \geq 0$ for each $i$; $x > 0$ means $x \geq 0$ and $x \neq 0$. 

Consider a CRN $\mathcal{R}$ involving $k$ chemical species $X_1, \ldots, X_k$ with corresponding concentration vector $x= (x_1, \ldots, x_k)^t$ and $l$ irreversible reactions. Fixing some ordering on the reactions one can define nonnegative $n \times m$ matrices $L$ and $R$ as follows: $L_{ij}$ (resp., $R_{ij}$) is the stoichiometry of species $X_i$ in the source complex (resp., target complex) of reaction $j$. The {\em irreversible stoichiometric matrix} of $\mathcal{R}$ is $\Gamma=R-L$. The $j$th column of $\Gamma$ is termed the {\em reaction vector} for the $j$th reaction.

If the reactions of $\mathcal{R}$ proceed with rates $v_1(x), v_2(x),\ldots, v_l(x)$, then the evolution of the species concentrations is governed by the ODE:
\begin{equation}
\label{genCRN}
\dot x = \Gamma v(x).
\end{equation}
where $v(x) =(v_1(x), v_2(x),\ldots, v_l(x))^t$ is the {\em rate function} of $\mathcal{R}$. The allowed choices of $v(x)$ (and even the domain of $v$) depend on various modelling choices; but many reasonable choices imply that $v(x) \gg 0$ for any $x \gg 0$. In this case we say that $\mathcal{R}$ has {\em positive kinetics}. It is also convenient to assume that $v(x)$ is locally Lipschitz on $\mathbb{R}^k_{\gg 0}$ in order to ensure a local flow on $\mathbb{R}^k_{\gg 0}$ associated with (\ref{genCRN}).

\section{Counting unlabelled CRNs: the basic classes}

The key computational tool for this project is the powerful graph isomorphism software NAUTY \cite{nauty}. While NAUTY does not allow direct operation on edge-labelled digraphs or multidigraphs, these can be represented as layered digraphs. In brief, individual vertices become sets of vertices, additional vertex colouring is introduced, and edges with different labels become edges between vertices of different colours as described in the section {\em Isomorphism of edge-coloured graphs} of the NAUTY documentation at \url{http://users.cecs.anu.edu.au/~bdm/nauty/nug26.pdf}. Via this process, a $k$-species, $l$-reaction 2-CRN can be represented as an ordinary digraph on $2(k+l)$ vertices: edges in the PN graph which were labelled $1$ or $2$ now correspond to arcs between different sets of vertices in the digraph. Here, this digraph is termed the {\em layered digraph} of the CRN, and two CRNs are isomorphic if and only if their corresponding layered digraphs are isomorphic. A 2-CRN has a layered digraph with {\em four} vertex colours, two corresponding to different layers of species vertices, and two corresponding to different layers of reaction vertices. (Indeed, an at most trimolecular CRN also has a 4-coloured layered digraph on $2(k+l)$ vertices, but if larger total stoichiometry is allowed, then additional vertices and colours need to be introduced.) 

\begin{remark}[Canonical labelling]
NAUTY can be used to canonically label digraphs while respecting the partition of the vertices, and so we can get a canonical fully labelled representative of a given unlabelled CRN, by first computing its layered digraph, applying NAUTY's canonical labelling, and then converting back to a CRN. Note that the canonical labelling gives both a species order and a reaction order, so that after canonical labelling comparison of two CRNs can amount to a simple string comparison. 
\end{remark}

\subsection{Counting irreversible 2-CRNs} 
\label{seccountbasic}
We first describe a naive approach to enumerating unlabelled 2-CRNs. Modifications which greatly increase the speed and efficiency of the process are described in Sections~\ref{secomit}~to~\ref{secinherit}. All unlabelled $k$-species, $l$-reaction 2-CRNs can be generated as follows: 
\begin{enumerate}
\item irreversible reactions are ordered pairs of distinct complexes: consequently there are a total of 
\[
n_R(k):=n_C(k)(n_C(k)-1)
\]
distinct irreversible reactions involving the $n_C$ 2-complexes; 
\item all possible sets of $l$ distinct reactions can be generated and stored as layered digraphs, represented in {\tt digraph6} format. There are
\[
N_{k,l} := {n_R(k) \choose l} = {{k+2 \choose 2}\left({k+2 \choose 2}-1\right)\choose l}
\]
of these CRNs. Note that $N_{k,l}$ counts the number of $k$-species, $l$-reaction, species-labelled 2-CRNs. For fixed $k$, and $l$ small compared to $n_R(k)$, $N_{k,l}$ grows quite rapidly. For example:
\[
N_{4,1} = 210, \quad N_{4,2} = 21,\!945, \quad N_{4,3} = 1.52 \times 10^6, \quad N_{4,4} = 7.87 \times 10^7, \quad \ldots
\]

\item the NAUTY program {\tt shortg} is used to canonically label and remove isomorphs from this list of CRNs, respecting the species-reaction bipartition. We are left with $\overline{N_{k,l}}$ canonically labelled CRNs. 
For example:
\[
\overline{N_{4,1}} = 22, \quad \overline{N_{4,2}} = 1,\!171, \quad \overline{N_{4,3}} = 67,\!257, \quad \overline{N_{4,4}} = 3.33 \times 10^6, \quad \ldots
\]
\end{enumerate}

\begin{remark}[Complete 2-CRNs]
A 2-CRN with $k$ species can have no more than $n_R(k)$ reactions. The unique 2-CRN with $k$ species and $n_R(k)$ reactions can be thought of as a ``complete'' 2-CRN with $k$ species, and it contains all other $k$-species 2-CRNs as subnetworks obtained by removing some reactions. 
\end{remark}

Table~\ref{tabdat} provides some data on the numbers $N_{k,l}$ and $\overline{N_{k,l}}$ of species-labelled and unlabelled CRNs defined above for CRNs with upto five species and five reactions. 

\begin{table}[h]
\centering
\begin{tabular}{cc|c|c|c|c|c|c} 
\cline{3-7} & & \multicolumn{5}{ c| } {number of reactions $l$} \\ 
\cline{3-7} & & 1 & 2 & 3 & 4 & 5\\ 
\cline{1-7} 
\multicolumn{1}{ |c| }{\parbox[t]{2mm}{\multirow{15}{*}{\rotatebox[origin=c]{90}{number of species $k$}}}} & \multicolumn{1}{  c| }{\multirow{3}{*}{1}} & $6$ & $15$ & $20$& $15$ & $6$\\
\multicolumn{1}{ |c|  }{}&&\cellcolor{black!20}$6$&\cellcolor{black!20}$15$&\cellcolor{black!20}$20$&\cellcolor{black!20}$15$&\cellcolor{black!20}$6$\\ 
\multicolumn{1}{ |c|  }{}&&$6$&$15$&$20$&$15$&$6$\\ 
\hhline{~*6-}
\multicolumn{1}{ |c|  }{}&\multicolumn{1}{  c| }{\multirow{3}{*}{2}} & $30$ & $435$ & $4,\!060$& $274,\!05$& $142,\!506$\\
\multicolumn{1}{ |c|  }{}&&\cellcolor{black!20}$16$&\cellcolor{black!20}$225$&\cellcolor{black!20}$2,\!044$&\cellcolor{black!20}$13,\!755$&\cellcolor{black!20}$71,\!344$\\ 
\multicolumn{1}{ |c|  }{}&&$15$&$218$&$2,\!030$&$13,\!703$&$71,\!253$\\ 
\cline{2-7}
\multicolumn{1}{ |c|  }{}&\multicolumn{1}{  c| }{\multirow{3}{*}{3}} & $90$ & $4,\!005$ & $117,\!480$& $2,\!555,\!190$& $43,\!949,\!268$\\
\multicolumn{1}{ |c|  }{}&&\cellcolor{black!20}$21$&\cellcolor{black!20}$720$&\cellcolor{black!20}$19,\!934$&\cellcolor{black!20}$427,\!770$&\cellcolor{black!20}$7,\!334,\!010$\\ 
\multicolumn{1}{ |c|  }{}&&$15$&$668$&$19,\!580$&$425,\!865$&$7,\!324,\!878$\\ 
\cline{2-7}
\multicolumn{1}{ |c|  }{}&\multicolumn{1}{  c| }{\multirow{3}{*}{4}} & $210$ & $21,\!945$ & $1,\!521,\!520$& $78,\!738,\!660$& $3,\!244,\!032,\!792$\\
\multicolumn{1}{ |c|  }{}&&\cellcolor{black!20}$22$&\cellcolor{black!20}$1,\!171$&\cellcolor{black!20}$67,\!257$&\cellcolor{black!20}$3,\!328,\!704$&\cellcolor{black!20}$135,\!622,\!844$\\ 
\multicolumn{1}{ |c|  }{}&&$9$&$915$&$63,\!397$&$3,\!280,\!778$&$135,\!168,\!033$\\ 
\cline{2-7}
\hhline{*{6}~|*2-|}
\multicolumn{1}{ |c|  }{}&\multicolumn{1}{  c| }{\multirow{3}{*}{5}} & $420$ & $87,\!990$ & $12,\!259,\!940$& $1,\!278,\!098,\!745$& $106,\!337,\!815,\!584$\\
\multicolumn{1}{ |c|  }{}&&\cellcolor{black!20}$22$&\cellcolor{black!20}$1,\!375$&\cellcolor{black!20}$122,\!939$&\cellcolor{black!20}$11,\!223,\!502$&\cellcolor{black!20}$899,\!358,\!555$\\ 
\multicolumn{1}{ |c|  }{}&&$4$&$734$&$102,\!167$&$10,\!650,\!823$&$886,\!148,\!464$\\ 

 \cline{1-7}
 \cline{2-7}
 \cline{4-7}\\
 \end{tabular}
\caption{\label{tabdat} Total number of $k$-species, $l$-reaction 2-CRNs for $k,l = 1, \ldots, 5$. In each box, the top entry is $N_{k,l}$, the total number of species-labelled $k$-species, $l$-reaction 2-CRNs; the middle entry (highlighted) is $\overline{N_{k,l}}$, the number of unlabelled $k$-species, $l$-reaction 2-CRNs, computed with the help of {\tt NAUTY}; and the bottom entry is $\lceil N_{k/l}/k!\rceil$, the lower bound on $\overline{N_{k,l}}$ as discussed in Section~\ref{secautomorph}. }
\end{table}

Note that for a fixed number of species $k$, $\lceil N_{k/l}/k!\rceil/\overline{N_{k,l}}$ gets close to $1$ as $l$ increases, reflecting the fact that the probability of symmetries in a randomly chosen CRN with many reactions compared to the number of species is small. For example, $N_{4,4}/\overline{N_{4,4}} \simeq 23.65$, namely, the average orbit size of the $4$-species, $4$-reaction CRNs under the action of $S_4$ is very close to $4!$. 

\subsection{Counting reversible 2-CRNs} All {\em reversible} unlabelled 2-CRNs with $k$ species and $l$ reversible reactions can be generated similarly to the irreversible case. A 2-CRN with $k$ species and $l$ reversible reactions is of course a CRN with $k$ species and $2l$ irreversible reactions. However the reversible 2-CRNs are enumerated directly as follows. 
\begin{enumerate}
\item Reversible reactions are {\em unordered} pairs of distinct complexes: consequently there are a total of $n_R^r(k) := {n_C(k) \choose 2}$ of these; 
\item all possible sets of $l$ distinct reversible reactions are generated and stored as layered digraphs.
There are
\[
N^r_{k,l} := {n^r_R(k) \choose l} = {{{k+2 \choose 2}\choose 2}\choose l}
\]
of these CRNs;
\item the NAUTY program {\tt shortg} is used to canonically label and remove isomorphs from this list, respecting the species-reaction bipartition. We are left with $\overline{N^r_{k,l}}$ canonically labelled CRNs. 
By comparison with the numbers above for irreversible reactions, for example:
\[
\overline{N^r_{4,1}} = 13, \quad \overline{N^r_{4,2}} = 325, \quad \overline{N^r_{4,3}} = 8,\!713, \quad \overline{N^r_{4,4}} = 205,\!948, \quad \ldots
\]
\end{enumerate}

\subsection{Omitting obvious isomorphs}
\label{secomit}
As we see from Table~\ref{tabdat}, the size of the enumeration problem grows rapidly with the number of species and reactions. For example, there are $N_{4,5} = 3,\!244,\!032,\!792$ species-labelled CRNs with four species and five reactions which fall into $\overline{N_{4,5}} = 135,\!622,\!844$ isomorphism classes. Handling the $N_{5,5} = 106,\!337,\!815,\!584$ species-labelled CRNs with five species and five reactions is a challenge: just storing these CRNs in (uncompressed) {\tt digraph6} format would take about 6 TB of space. On the other hand parsing the data several times would lead to a large increase in terms of computation time.  

An easy observation considerably reduces the challenge by allowing us to omit many isomorphs of a CRN at the point where the CRNs are generated. Consider the set of 2-CRNs on $k$ species $X_1, \ldots, X_k$. Each unlabelled 2-CRN on $k$ species has at least one species-labelled representative satisfying 
\[
\mathrm{deg}\,X_1 \geq \mathrm{deg}\,X_2 \geq \cdots \geq \mathrm{deg}\,X_k
\]
where $\mathrm{deg}\,X_i$ refers to the degree of the vertex corresponding to species $X_i$ in the PN graph of the CRN. We may thus insist that $\mathrm{deg}\,X_1 \geq \cdots \geq \mathrm{deg}\,X_k$ when generating CRNs in step (2) above; this considerably cuts down both on the size of storage needed to hold CRNs for isomorphism testing, and the total number of relabellings required. 

\subsection{Using invariants}
\label{secinv}

It is natural -- and indeed necessary -- to divide up the raw species-labelled CRNs using isomorphism invariants which are easily computed before attempting to remove isomorphs. Thus one might, for example, count how many times the stoichiometry $2$ figures in a source complex in a CRN, or a target complex of the CRN, and/or compute the total stoichiometry of all species in all reactions in the CRN. One can then use these invariants to construct multiple independent lists of $k$-species, $l$-reaction 2-CRNs, before using {\tt shortg} to remove isomorphs from each list, and finally merging the lists. The larger sets of 2-CRNs at \url{https://reaction-networks.net/networks/} were enumerated using this approach, combined with the removal of obvious isomorphs discussed in the previous section.

\subsection{Enumeration by inheritance}
\label{secinherit}

In the light of the explosion in problem size, an alternative to enumerating species-labelled CRNs and then separating these into isomorphism classes is to build larger CRNs from smaller ones. For example, given representatives of unlabelled CRNs with $k$ species and $l$ reactions, we may hope to find representatives of the unlabelled 2-CRNs with $k$ species and $l+1$ reactions as follows: we take each representative 2-CRN with $k$ species and $l$ reactions, and add to it every possible reaction on $k$ species which does not already occur in it; we then remove isomorphs from this list of $k$-species, $(l+1)$-reaction 2-CRNs. This certainly gives a complete list of nonisomorphic $k$-species, $(l+1)$-reaction 2-CRNs, since the removal of any reaction from a $k$-species, $(l+1)$-reaction 2-CRN leaves a valid $k$-species, $l$-reaction 2-CRN.

Note that we cannot easily interchange species and reactions in this approach and build nonisomorphic CRNs by adding in species to smaller CRNs. The reason is that there are valid CRNs with $k+1$ species and $l$ reactions, which do not leave a valid CRN after removal of any species. As an example consider the CRN 
\[
A+B\rightarrow A, \,\,B \rightarrow A+B.
\]
Removal of $A$ leaves $B \rightarrow 0$, $B \rightarrow B$, while removal of $B$ leaves $A\rightarrow A$, $0 \rightarrow A$, neither of which is a valid CRN (recall that in our definition of a CRN, source and target complexes of a reaction are distinct). Thus this CRN has no induced subCRNs involving $1$ species and $2$ reactions, and so cannot be built by adding species into a CRN involving $1$ species and $2$ reactions.

It is not clear to what extent an inheritance-based approach reduces the size of the basic enumeration problem. For example, consider the $3,\!328,\!704$ unlabelled $4$-species $4$-reaction $2$-CRNs. As there are 210 at most bimolecular $4$-species reactions, finding the $4$-species $5$-reaction $2$-CRNs by this approach involves testing approximately $3,\!328,\!704 \times 210 \simeq 7 \times 10^8$ CRNs. Although this is somewhat less than $N_{4,5} \simeq 3.2 \times 10^9$, the combination of approaches in Sections~\ref{secomit}~and~\ref{secinv} proves more efficient at enumerating the $4$-species $5$-reaction $2$-CRNs than an inheritance based approach. However, the inheritance-based approach is useful for a number of problems where we want specifically to enumerate CRNs with a given induced subgraph.

\section{Interesting subclasses of CRNs}

The raw CRNs enumerated by the techniques described in the previous section may be interesting from a purely combinatorial point of view, but we may wish to exclude some of them for various reasons. Below is a (far from exhaustive) list of some interesting subclasses of CRNs with some comments on their enumeration. Relationships among the classes are summarised diagrammatically in Section~\ref{secrel}. 

\subsection{Genuine CRNs} The definition of a CRN does not exclude the possibility that some species participate in no reactions. However, when analysing CRNs with $k$ species and $l$ reactions one may want to exclude ones with unused species (corresponding to isolated species vertices in the PN graph). CRNs which do not have such unused species are termed {\em genuine} (for want of a better word).

\begin{remark}[Isolated reaction vertices are already ruled out]
Note that the PN graph of a CRN cannot have an isolated reaction vertex as this would correspond to the reaction $0 \rightarrow 0$ which is not a valid reaction.
\end{remark}

\begin{remark}[Maximum number of species in a genuine CRN]
Since an at most bimolecular reaction involves a maximum of $4$ species, a genuine 2-CRN with $l$ reactions can have no more than $4l$ species. 
\end{remark}

\begin{remark}[The number of genuine CRNs]
Let $\overline{N^G_{k,l}}$ refer to the number of unlabelled genuine 2-CRNs with $k$ species and $l$ reactions. We have the formula:
\[
\overline{N^G_{k,l}} = \overline{N_{k,l}} - \overline{N_{k-1, l}}\,.
\]
This follows because the $k$-species, $l$-reaction CRNs which are {\em not} genuine are exactly those obtained from $(k-1)$-species, $l$-reaction CRNs via addition of a redundant species. 
\end{remark}

\subsection{Indecomposable CRNs} Rather than just avoiding CRNs with unused species, one may wish to exclude CRNs with disconnected PN graphs (namely CRNs whose species can be divided into two nonempty non-interacting subsets). We refer to CRNs with disconnected PN graphs as {\em decomposable}, while CRNs with connected PN graphs are {\em indecomposable}. If we are interested in searching for new dynamical behaviours which arise in larger CRNs (under any reasonable modelling assumptions), then exluding decomposable CRNs is natural, as their dynamics decouples into that of the smaller CRNs of which they are composed. 

\begin{remark}[Indecomposable CRNs are genuine]
The indecomposable CRNs are clearly a subset of the genuine CRNs. 
\end{remark}

\begin{remark}[Testing for indecomposability]
We may test whether the PN graph of a CRN is connected with a simple depth-first search beginning with any vertex of the PN graph where we can traverse arcs in either direction.
\end{remark}

\subsection{Dynamically nontrivial CRNs} We might wish to exclude CRNs which are in some way uninteresting from a dynamical point of view. Consider a CRN $\mathcal{R}$ consisting of $k$ species and $l$ irreversible reactions with $k \times l$ stoichiometric matrix $\Gamma$. $\mathcal{R}$ is referred to as {\em dynamically trivial} if there exists a linear scalar function which increases along all positive orbits for any positive kinetics (see Section~\ref{secode}), and {\em dynamically nontrivial} otherwise. Equivalently, $\mathcal{R}$ is dynamically trivial if there exists a vector $q>0$ in $\mathrm{im}\,\Gamma^{\mathrm{t}}$. To see the equivalence, note that if there exists $p$ s.t. $\Gamma^{\mathrm{t}}p = q > 0$, then for any kinetics such that $x \gg 0 \Rightarrow v(x) \gg 0$, we have
\[
\frac{\mathrm{d}}{\mathrm{d}t}p^{\mathrm{t}}x = p^{\mathrm{t}}\dot x = q^{\mathrm{t}}v(x) >0
\]
and thus $p^{\mathrm{t}}x$ increases along orbits at every point in $\mathbb{R}^k_{\gg 0}$. 

By standard arguments (Remark~\ref{remDN}), a dynamically trivial CRN $\mathcal{R}$ can have no limit sets intersecting $\mathbb{R}^k_{\gg 0}$. If we are primarily interested in CRNs which potentially admit positive equilibria, periodic orbits, chaos, etc., then we would wish immediately to exclude the dynamically trivial CRNs. 

\begin{remark}[Dynamically nontrivial CRNs have no positive limit sets]
\label{remDN}
For completeness, we sketch the proof of this assertion. Let $\phi$ be the local flow on $\mathbb{R}^k_{\gg 0}$ associated with $\dot x = \Gamma v(x)$ and suppose that there exists some $y \gg 0$ in the limit set of $\phi$. In other words, there exists some $x \in \mathbb{R}^k_{\gg 0}$ and a sequence of times $t_k \to \infty$ as $k \to \infty$ s.t. 
\[
\lim_{k \to \infty}\phi_{t_k}(x) = y\,.
\]
Set $H(x) = p^\mathrm{t}x$. Then, by continuity of $H$, $\lim_{k \to \infty}H(\phi_{t_k}(x)) = H(y)$ and since $\dot H(x) > 0$, $H(\phi_{t_k}(x)) < H(y)$ and consequently $H(\phi_{t}(x)) < H(y)$ for all $t > 0$. On the other hand, again since $\dot H(x) > 0$, $H(\phi_s(y)) > H(y)$ for each $s > 0$. As $\phi_{t_k}(x) \to y$ as $k \to \infty$, by continuity of $H$, for sufficiently large $k$, $H(\phi_s(\phi_{t_k}(x))) = H(\phi_{s+t_k}(x)) > H(y)$ which contradicts the assertion that $H(\phi_{t}(x)) < H(y)$ for all $t>0$.
\end{remark}

\begin{remark} To state that a CRN is dynamically nontrivial only implies the nonexistence of an increasing linear functional; we do not claim that a dynamically nontrivial CRN {\em must} admit some limit set intersecting the positive orthant for arbitrary positive kinetics. Note also that a dynamically trivial CRN may have nontrivial behaviour on the boundary of the positive orthant. 
\end{remark}

\begin{remark}[Testing whether a given CRN is dynamically trivial]This is a feasibility problem, which can be solved, for example, with the help of the linear programming package {\tt GLPK} (\url{http://www.gnu.org/software/glpk/glpk.html}).
\end{remark}

\subsection{Weakly reversible CRNs} Another class of CRNs which are of interest are the {\em weakly reversible} CRNs \cite{hornjackson}. A CRN is weakly reversible if every connected component (CC) of its complex graph is a strongly connected component (SCC). Weakly reversible CRNs are rare among general CRNs. For example, only $11,\!544$ out of the $135,\!622,\!844$ unlabelled $4$-species, $5$-reaction 2-CRNs are weakly reversible. This is about $0.009\%$ of the total. And in fact only $6,\!552$ of these are genuine. However, given their importance in classical CRN theory \cite{feinberg}, the enumeration of weakly reversible CRNs is worthwhile.

\begin{remark}[Weakly reversible CRNs are dynamically nontrivial] 
\label{remWR}
Let $\Gamma$ be the irreversible stoichiometric matrix of a CRN $\mathcal{R}$. As is well known and easily proved, if $\mathcal{R}$ is weakly reversible, then $\mathrm{ker}\,\Gamma$ includes a positive vector. Consequently, $\mathrm{im}\,\Gamma^t$ includes no vector $>0$ (Theorem~3' in \cite{benisrael}, for example) and the claim follows.
\end{remark}

\begin{remark}[Reversible CRNs are weakly reversible]
This is immediate from the definitions. 
\end{remark}

\begin{remark}[Testing for weak reversibility]
The test for whether a CRN $\mathcal{R}$ is weakly reversible is a standard graph theoretic test on the complex graph of $\mathcal{R}$. We can, for example, take each CC of the complex graph and check if it is an SCC using Tarjan's algorithm \cite{Tarjan}. (Note that unlike the test for indecomposability, we need to examine the complex graph, not the PN graph.)
\end{remark}

\subsection{Fully open CRNs}

An important and highly studied subclass of CRNs are the ``fully open CRNs''. One possible interpretation of ``fully open'' would be to class a CRN with stoichiometric matrix $\Gamma$ as fully open if $\mathrm{rank}\,\Gamma$ (namely, the dimension of the stoichiometric subspace, $\mathrm{im}\,\Gamma$) is equal to the total number of species. Equivalently, the system has no linear first integrals. However, for some purposes (see, for example, \cite{banajipanteaMPNE}) it is useful to adopt a stricter notion: a CRN involving species $\mathrm{X}_1, \ldots, \mathrm{X}_k$ is defined to be {\em fully open} if and only if it includes all the reactions $0 \rightleftharpoons \mathrm{X}_i$ ($i = 1, \ldots, k$). This is the notion adopted here. Reactions not of the form $0 \rightarrow \mathrm{X}_i$ or $\mathrm{X}_i \rightarrow 0$ are termed {\em non-flow reactions}. 

\begin{remark}[Fully open CRNs are genuine and dynamically nontrivial]
Clearly, a fully open CRN is genuine as each species participates in the reactions $0 \rightarrow \mathrm{X}_i$ and $\mathrm{X}_i \rightarrow 0$. It is also easily seen that a fully open CRN is dynamically nontrivial: the irreversible stoichiometric matrix of a fully open CRN includes a positive vector in its kernel and hence fully open CRNs are dynamically nontrivial by the reasoning in Remark~\ref{remWR}. 
\end{remark}

One way of enumerating fully open 2-CRNs with $k$ reactions and $l$ non-flow reactions is to consider all $k$-species, $l$-reaction 2-CRNs and remove any which include a reaction of the form $0 \rightarrow \mathrm{X}_i$ or $\mathrm{X}_i \rightarrow 0$. Noting that two fully open CRNs are isomorphic if and only if they are isomorphic after removal of the flow reactions $0 \rightleftharpoons \mathrm{X}_i$, these are now precisely the fully open 2-CRNs with all the reactions $0 \rightleftharpoons \mathrm{X}_i$ removed. 

Alternatively, fully open 2-CRNs may be enumerated directly. This proceeds in a similar fashion to enumerating general 2-CRNs. We describe the process for the general case; the special case of fully open, reversible, 2-CRNs is easily obtained via minor modifications. 
\begin{enumerate}
\item From the total of $n_C(k)(n_C(k)-1)$ distinct irreversible reactions involving all 2-complexes we exclude reactions of the form $0 \rightarrow \mathrm{X}_i$ and $\mathrm{X}_i \rightarrow 0$ leaving $n_{R,o}(k):=n_C(k)(n_C(k)-1) - 2k$ distinct non-flow reactions;
\item all possible sets of $l$ distinct non-flow reactions are enumerated. There are
\[
N^o_{k,l}:={n_{R,o}(k) \choose l} = {{k+2 \choose 2}\left({k+2 \choose 2}-1\right) - 2k \choose l}
\]
of these; 
\item the NAUTY program {\tt shortg} is used to canonically label and remove isomorphs from this list of CRNs, respecting the species-reaction bipartition. 
\item The reactions $0 \rightarrow \mathrm{X}_i$ and $\mathrm{X}_i \rightarrow 0$ are added back into the canonically labelled CRNs. 
\end{enumerate}

\begin{remark}
In order to save on space, the fully open 2-CRNs stored at \url{https://reaction-networks.net/networks/} have the reactions $0 \rightarrow \mathrm{X}_i$ and $\mathrm{X}_i \rightarrow 0$ removed. Thus to reconstruct them from the stored networks, these need to be added in. 
\end{remark}

The strategies to speed up enumeration in Sections~\ref{secomit}~to~\ref{secinherit} apply equally to fully open 2-CRNs. Although the numbers $N^o_{k,l}$ grow almost as fast as the numbers $N_{k,l}$, a fully open 2-CRN with $k$ species and $l$ non-flow reactions is actually a 2-CRN with $k$ species and $l+2k$ reactions and is automatically dynamically nontrivial; consequently a much greater proportion of fully open 2-CRNs with $k$ species and $l$ non-flow reactions are likely to be dynamically interesting than the corresponding proportion for 2-CRNs with $k$ species and $l$ reactions.

%
%

\section{Relationships among classes of CRNs discussed}
\label{secrel}

Inclusions amongst the various classes of CRNs discussed here are illustrated in the diagram below. $A \rightarrow B$ means $A \supseteq B$, and in fact all the inclusions are strict. The following acronyms are used: DN = dynamically nontrivial, WR = weakly reversible, FO = fully open, R = reversible, G = genuine, I = indecomposable.

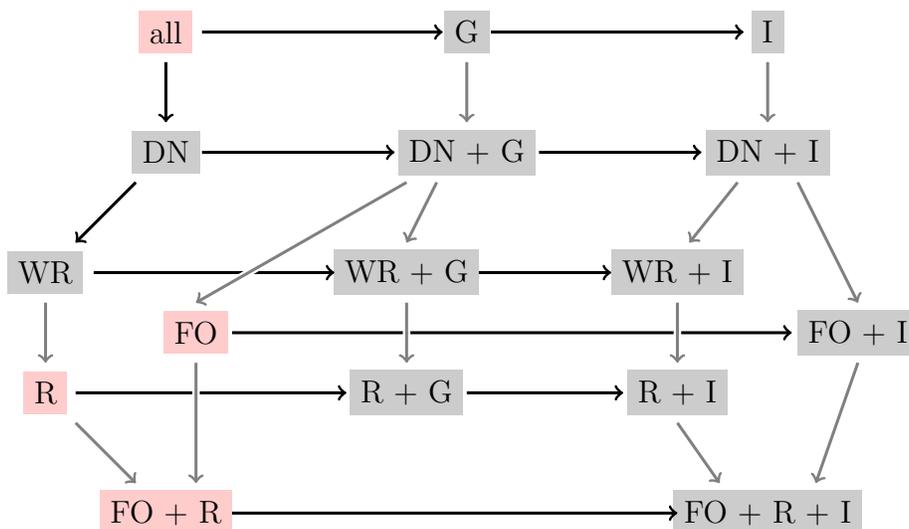
\begin{center}
\begin{tikzpicture}[scale=0.8]

\node[shape=rectangle,fill=red!20] at (0,10) {all};
\node[shape=rectangle,fill=black!20] at (0,8) {DN};
\node[shape=rectangle,fill=black!20] at (-2,6) {WR};
\node[shape=rectangle,fill=red!20] at (0.5,5) {FO};
\node[shape=rectangle,fill=red!20] at (-2,4) {R};
\node[shape=rectangle,fill=red!20] at (0,2) {FO + R};
\node[shape=rectangle,fill=black!20] at (5, 10) {G};
\node[shape=rectangle,fill=black!20] at (5, 8) {DN + G};
\node[shape=rectangle,fill=black!20] at (4, 6) {WR + G};
\node[shape=rectangle,fill=black!20] at (4, 4) {R + G};
\node[shape=rectangle,fill=black!20] at (10, 10) {I};
\node[shape=rectangle,fill=black!20] at (10, 8) {DN + I};
\node[shape=rectangle,fill=black!20] at (8.5, 6) {WR + I};
\node[shape=rectangle,fill=black!20] at (8.5, 4) {R + I};
\node[shape=rectangle,fill=black!20] at (11.5, 5) {FO + I};
\node[shape=rectangle,fill=black!20] at (10, 2) {FO + R + I};


\draw [->, line width=0.04cm] (0.6,10) -- (4.6,10);
\draw [->, line width=0.04cm] (5.4,10) -- (9.6,10);
\draw [->, line width=0.04cm] (0.6,8) -- (3.8,8);
\draw [->, line width=0.04cm] (6.2,8) -- (8.9,8);
\draw [->, line width=0.04cm] (-1.2,6) -- (2.8,6);
\draw [->, line width=0.04cm] (5.2,6) -- (7.4,6);
\draw [->, line width=0.04cm] (1.1,5) -- (10.4,5);
\draw [->, line width=0.04cm] (-1.5,4) -- (3,4);
\draw [->, line width=0.04cm] (5,4) -- (7.6,4);
\draw [->, line width=0.04cm] (1.1,2) -- (8.5,2);


\draw [->, line width=0.04cm] (0,9.5) -- (0,8.5);
\draw [->, line width=0.04cm] (-0.5,7.5) -- (-1.5,6.5);
\draw [->, line width=0.1cm,color=white] (-2,5.5) -- (-2,4.5);
\draw [->, color=black!50, line width=0.04cm] (-2,5.5) -- (-2,4.5);
\draw [->, color=black!50, line width=0.04cm] (-1.5,3.5) -- (-0.5,2.5);
\draw [->, color=white, line width=0.1cm] (0.5,4.5) -- (0.5,2.5);
\draw [->, color=black!50, line width=0.04cm] (0.5,4.5) -- (0.5,2.5);


\draw [->,color=black!50,  line width=0.04cm] (5,9.5) -- (5,8.5);
\draw [->, color=black!50,  line width=0.04cm] (4.5,7.5) -- (4,6.5);
\draw [->, line width=0.1cm,color=white] (4,5.5) -- (4,4.5);
\draw [->, color=black!50, line width=0.04cm] (4,5.5) -- (4,4.5);

\draw [->,color=white,  line width=0.1cm] (4,7.5) -- (0.5,5.5);
\draw [->,color=black!50,  line width=0.04cm] (4,7.5) -- (0.5,5.5);


\draw [->, color=black!50, line width=0.04cm] (10,9.5) -- (10,8.5);
\draw [->, color=black!50, line width=0.04cm] (9.5,7.5) -- (8.7,6.5);
\draw [->, color=black!50, line width=0.1cm, color=white] (10.5,7.5) -- (11.5,5.5);
\draw [->, color=black!50, line width=0.04cm] (10.5,7.5) -- (11.5,5.5);
\draw [->, color=white, line width=0.1cm] (8.5,5.5) -- (8.5,4.5);
\draw [->, color=black!50, line width=0.04cm] (8.5,5.5) -- (8.5,4.5);
\draw [->, color=black!50, line width=0.04cm] (8.5,3.5) -- (9.2,2.5);
\draw [->, color=black!50, line width=0.1cm,color=white] (11.5,4.5) -- (10.8,2.5);
\draw [->, color=black!50, line width=0.04cm] (11.5,4.5) -- (10.8,2.5);

\end{tikzpicture}
\end{center}

The sets ``all'', ``FO'' (fully open), ``R'' (reversible) and ``FO + R'' (fully open, reversible), highlighted in red, are directly enumerated as described above. The remaining sets are enumerated by taking some parent set (connected to the set via a bold arrow) and checking for additional properties. For example, the dynamically nontrivial CRNs are obtained from the set of all CRNs by testing each CRN for the property of being dynamically nontrivial; the weakly reversible CRNs are obtained from the set of all dynamically nontrivial CRNs by testing each for the property of weak reversibility; the dynamically nontrivial, genuine CRNs are obtained from the dynamically nontrivial CRNs by extracting those CRNs without isolated species. And so forth.

\bibliographystyle{unsrt}

\end{document}